\documentclass{article}
\def\Ref#1{(\ref{#1})}
\usepackage{amsmath}
\begin{document}
\begin{titlepage}
\begin{center}
{\large \bf Phase transitions in autonomous reaction--diffusion
systems on a one--dimensional lattice with boundaries} \vskip
2\baselineskip \centerline {\bf
 Amir Aghamohammadi $^{a}$ \footnote {e-mail:mohamadi@theory.ipm.ac.ir}
 {\rm and}
 Mohammad Khorrami $^{b}$ \footnote {e-mail:mamwad@iasbs.ac.ir}}
 \vskip 2\baselineskip
{\it $^a$ Department of Physics, Alzahra University, Tehran 19384, IRAN}\\
{\it $^b$ Institute for Advanced Studies in Basic Sciences,}
\\ {\it P. O. Box 159, Zanjan 45195, IRAN}\\
\end{center}
\vskip 2cm
{\bf PACS numbers:} 82.20.Mj, 02.50.Ga, 05.40.+j

\noindent{\bf Keywords:} reaction--diffusion, phase transition

\begin{abstract}
\noindent The family of {\it autonomous} reaction--diffusion
models on a one--dimensional lattice with boundaries is studied.
By autonomous, it is meant that the evolution equation for
$n$--point functions contain only $n$- or less- point functions.
It is shown that these models exhibit a static and a dynamic phase
transition.
\end{abstract}
\end{titlepage}
\newpage
\section{Introduction}
In recent years, reaction--diffusion systems have been studied by
many people. As mean--field techniques, generally, do not give
correct results for low--dimensional systems, people are motivated
to study exactly--solvable stochastic models in low dimensions.
Moreover, solving one--dimensional systems should in principle be
easier. Exact results for some models on a one--dimensional
lattice have been obtained, for example in [1--12]. Different
methods have been used to study these models, including analytical
and asymptotic methods, mean field methods, and large--scale
numerical methods.

Some interesting problems in non--equilibrium systems are
non--equilibrium phase transitions described by phenomenological
rate equations, and the way the system relaxes to its steady
state. Kinetic generalizations of the Ising model, for example the
Glauber model or the Kawasaki model, are such phenomenological
models and have been studied extensively [13--18]. Combination of
the Glauber and the Kawasaki dynamics has been also considered
[19--22].

In \cite{MA}, an asymmetric generalization of the
zero--temperature Glauber model on a lattice with boundaries was
introduced. It was shown there that in the thermodynamic limit,
the system shows two kinds of phase transitions. One of these is a
static phase transition, the other a dynamic one. The static phase
transition is controlled by the reaction rates, and is a
discontinuous change of the behavior of the derivative of the
stationary particle density at the end points, with respect to the
reaction rates. The dynamic phase transition is controlled by the
injection- and extraction- rates of the particles at the end
points, and is a discontinuous change of the relaxation time
towards the stationary configuration. Other generalizations of the
Glauber model consist of, for example, alternating--isotopic
chains and alternating--bound chains (see \cite{GO}, for example).
People have also considered phase transitions induced through
boundary conditions (see \cite{HS,RIK}, for example).

In \cite{Sc}, a ten--parameter family of reaction--diffusion
reactions was introduced for which the evolution equation of
$n$--point functions contain only $n$- or less- point functions.
What we do in this paper, is to investigate these systems on a
finite lattice with boundaries. It will be shown that the
stationary behavior of the system is effectively controlled by two
parameters. On the one--dimensional boundary of this
two--dimensional parameter space, there exists a phase transition
(in the thermodynamic limit, when the lattice becomes infinite),
we call which a static phase transition.

The relaxation time toward the stationary state of the system may
depend on the injection- and extraction- rates at each of the
boundaries. It will be shown that in the thermodynamic limit there
are three regions, in one of them this time is independent of the
injection- and extraction- rates, in the second it depends on the
injection- and extraction rates at on end, and in the third it
depends on the injection- and extraction rates at the other end.
This is called the dynamic phase transition.

The scheme of the paper is as follows. In section 2, autonomous
reaction--diffusion systems with boundaries are introduced. In
section 3, the static phase transition of these systems is
investigated. Finally, in section 4, the dynamic phase transition
is studied.
\section{Autonomous reaction--diffusion systems with boundaries}
Consider a collection of particles drifting and reacting on a
one--dimensional lattice with $L$ sites. Each site may be
occupied, $A$, the state corresponding to which is denoted by
$|1\rangle$, or empty, $\emptyset$, the state of which is denoted
by $|0\rangle$. The rate of change of the state
$|\alpha\beta\rangle$ to the state $|\gamma\delta\rangle$ is
$H^{\gamma\delta}_{\alpha\beta}$. It is shown in \cite{Sc} that
the evolution equations for $n$--point functions are closed
(involve only $n$-- or less than $n$--point functions) iff the
following conditions are satisfied by $H$:
\begin{align}\label{1}
-H^{01}_{11}-H^{00}_{11}+H^{01}_{10}+H^{00}_{10}-H^{11}_{01}
-H^{10}_{01}+H^{11}_{00}+H^{10}_{00} &=:0,\nonumber\\
-H^{10}_{11}-H^{00}_{11}-H^{11}_{10}-H^{01}_{10}+H^{10}_{01}
+H^{00}_{01}+H^{11}_{00}+H^{01}_{00} &=:0.
\end{align}
Defining
\begin{align}\label{3}
u&:=H^{10}_{01}+H^{00}_{01}\nonumber\\
v&:=H^{01}_{10}+H^{00}_{10}\nonumber\\
\bar u&:=H^{10}_{11}+H^{00}_{11}\nonumber\\
\bar v&:=H^{01}_{11}+H^{00}_{11}\nonumber\\
w&:=H^{11}_{00}+H^{10}_{00}\nonumber\\
s&:=H^{11}_{00}+H^{01}_{00}\nonumber\\
\bar w&:=H^{11}_{01}+H^{10}_{01}\nonumber\\
\bar s&:=H^{11}_{10}+H^{01}_{10},
\end{align}
one can write \Ref{1} as
\begin{align}\label{3a}
u+s&=\bar u+\bar s\nonumber\\
v+w&=\bar v+\bar w.
\end{align}
At the end sites $1$ and $L$, there are also injection-- and
extraction--rates. The injection-- and extraction--rates in the
first site are denoted by $a$ and $a'$, respectively. The
corresponding rates in the last site are denoted by $b$ and $b'$.
It is then seen that
\begin{align}\label{2}
\langle\dot n_k\rangle =& -(v+w+u+s)\langle n_k\rangle + (v-\bar
v) \langle n_{k+1}\rangle +(u-\bar u)\langle
n_{k-1}\rangle\nonumber\\
& +w+s,\qquad 1<k<L \nonumber\\
\langle\dot n_1\rangle =& -(v+w)\langle n_1\rangle + (v-\bar v)
\langle n_2\rangle +w+a(1-\langle n_1\rangle )-a' \langle
n_1\rangle\nonumber\\
\langle\dot n_L\rangle =& -(u+s)\langle n_L\rangle + (u-\bar u)
\langle n_{L-1}\rangle +s+b(1-\langle n_L\rangle )-b'\langle
n_L\rangle ,
\end{align}
where $\langle n_k\rangle$ is the probability that the $k$-th site
is occupied. Comparing this with \cite{MA}, it is seen that the
model considered there is a special case of this model with $\bar
u=\bar v=w=s=0$.

\section{The static phase transition of the system}
The steady--state solution to (\ref{2}) is
\begin{equation}\label{4}
\langle n_k\rangle =C+D_1z_1^k+D_2z_2^{k-L-1},
\end{equation}
where $z_{1,2}$ satisfy
\begin{equation}\label{5}
-(u+v+w+s)+(v-\bar v)z_{1,2}+(u-\bar u)z^{-1}_{1,2}=0,
\end{equation}
and $z_2$ is that root the absolute value of which is greater. Let
us consider this equation more carefully. Defining three new
parameters $p$, $q$, and $r$ through
\begin{align}\label{6}
p:=&v-\bar v\nonumber\\
q:=&u-\bar u\nonumber\\
r:=&u+s+v+w=u+s+\bar v+\bar w\nonumber\\
=&\bar u+\bar s+v+w=\bar u+\bar s+\bar v+\bar w,
\end{align}
(where \Ref{3a} has been used) (\ref{5}) is rewritten as
\begin{equation}\label{7}
p\; z^2-r\; z+q=0.
\end{equation}
Using (\ref{6}) and the fact that the rates are nonnegative, it is
seen that
\begin{equation}\label{8}
r\geq |p|,|q|,|p+q|,|p-q|.
\end{equation}
The boundaries of the physical parameter space are thus
\begin{equation}\label{b1}
r=|p+q|, \hbox{ and }r=|p-q|.
\end{equation}
For $r=p+q$, it is seen that $\bar u=\bar v=s=w=0$, which means
that $AA$ and $\emptyset\emptyset$ don't change. So, there are two
equilibrium states on an infinite lattice without injection or
extraction; either all of the sites are occupied, or all of them
are unoccupied. For $p+q=-r$, one has $u=v=\bar s=\bar w=0$, which
means that $A\emptyset$ and $\emptyset A$ don't change. So, there
are two equilibrium states on an infinite lattice without
injection or extraction; $\cdots A\emptyset A\emptyset\cdots$ and
$\cdots \emptyset A\emptyset A\cdots$.

For $r=q-p$, one has $\bar u=v=s=\bar w=0$. The only nonzero rates
are then $H_{01}^{00}=H_{10}^{11}$ and $H_{11}^{01}=H_{00}^{10}$.
As all of the configurations can be converted to each other
through the reactions, the equilibrium state of the infinite
lattice without injection or extraction is unique. It is not
difficult to see that for the special case that these four nonzero
rates are equal, this state is
$\cdots(1/2)(|0\rangle+|1\rangle)\otimes(1/2)
(|0\rangle+|1\rangle)\otimes\cdots$. For $r=p-q$, one has $u=\bar
v=\bar s=w=0$. The only nonzero terms are then
$H_{10}^{00}=H_{00}^{11}$ and $H_{11}^{10}=H_{00}^{01}$. It is the
same as the case $r=q-p$, with left and right interchanged.

Also, if $r=0$, then $u=\bar u=v=\bar v=w=s=0$, so that
$\langle\dot n_k\rangle =0,\quad 1<k<L$. Neglecting this trivial
case, it is seen that $r$ is positive and there are two parameters
determining the behavior of the roots of (\ref{7}):
\begin{equation}\label{9}
P(z):=p'\; z^2-z+q'=0,
\end{equation}
where
\begin{align}\label{10}
p'&:=p/r\nonumber\\
q'&:=q/r.
\end{align}
Noting that $P(1)<0$ and $P(-1)>0$ for $r>|p+q|$, it is seen that
both of the roots of (\ref{9}) are real, one of them is between 1
and $-1$, the other is out of this interval. So, $|z_1|<1<|z_2|$
for $r>|p+q|$. In the thermodynamic limit $L\to\infty$,
\begin{align}\label{11}
\langle n_k\rangle &\approx C+D_1z_1^k,  &k\ll L\nonumber\\
\langle n_k\rangle &\approx C+D_2z_2^{k-L-1},  &L-k\ll L.
\end{align}
$z_1$, $z_2$, $C$, $D_1$, and $D_2$ are continuous functions of
the rates. So the behavior of $\langle n_k\rangle$ near the ends
of the lattice varies continuously with rates, and there is no
phase transition.

If $r=p+q$, one of the roots of (\ref{9}) is one, the other is
$q'/p'=q/p$. If $q>p$, then $z_1=1$ and $\langle n_k\rangle$ is
flat for $k\ll L$. This is independent of $p$ and $q$. However, if
$q<p$, then $z_2=1$ and the slope of $\langle n_k\rangle$ depends
on the rates. For $L-k\ll L$, a reverse behavior occurs. If $q<p$,
then $z_2=1$ and $\langle n_k\rangle$ is flat for $L-k\ll L$. If,
$q>p$, then $z_2=q/p$ and $\langle n_k\rangle$ varies with $k$ for
$L-k\ll L$. To summarize, one defines two effective roots $z_{\rm
l}$ and $z_{\rm r}$ for sites near $k=1$ (the left end) and $k=L$
(the right end), respectively. We then have
\begin{equation}\label{12}
z_{\rm l}=
\begin{cases}
1,&q>p\\ q/p,&q<p
\end{cases}
\end{equation}
and
\begin{equation}\label{13}
z_{\rm r}=
\begin{cases}
q/p,&q>p\\ 1,&q<p
\end{cases}
\end{equation}
So there is a phase transition at $p=q=r/2$. This corresponds to
$u=v=\bar s=\bar w$.

If $r=-p-q$, one of the roots of (\ref{9}) is $-1$ and the other
is $-q/p$. The same behavior is repeated, that is
\begin{equation}\label{14}
z_{\rm l}=
\begin{cases}
-1,&-q>-p\\ -q/p,&-q<-p
\end{cases}
\end{equation}
and
\begin{equation}\label{15}
z_{\rm r}=
\begin{cases}
-q/p,&-q>-p\\ -1,&-q<-p
\end{cases}
\end{equation}
Again, there is a phase transition at $p=q=-r/2$. This corresponds
to $\bar u=\bar v=s=w$.

It was seen that on two segments of the boundary of the physical
parameter space, there exists a static phase transition. These
segments ($r=|p+q|$) correspond to cases where the equilibrium
state on an infinite lattice without injection and extraction is
not unique. On the other segments of the boundary of the physical
parameter space ($r=|p-q|$), where the equilibrium state is unique
on an infinite lattice without injection and extraction, there is
no static phase transition for the lattice with boundaries.

\section{The dynamic phase transition of the system}
The homogeneous part of (\ref{2}) can be written as
\begin{equation}\label{16}
\langle\dot n_k\rangle =h_k^l\langle n_l\rangle .
\end{equation}
The eigenvalues and eigenvectors of the operator $h$ satisfy
\begin{align}\label{17}
E\; x_k&= -(v+w+u+s)x_k+(v-\bar v)x_{k+1}+(u-\bar u)x_{k-1},\quad
k\ne 1,L\nonumber\\
E\; x_1&= -(v+w+a+a')x_1+(v-\bar v)x_2,\nonumber\\
E\; x_L&= -(u+s+b+b')x_L+(u-\bar u)x_{L-1},
\end{align}
where the eigenvalue and the eigenvector have been denoted by $E$
and $x$, respectively. The solution to these is
\begin{equation}\label{18}
x_k=\alpha z_1^k+\beta z_2^k,
\end{equation}
where $z_j$'s satisfy
\begin{equation}\label{19}
E=-(v+w+u+s)+(v-\bar v)z+(u-\bar u)z^{-1},
\end{equation}
and
\begin{align}\label{20}
(v-\bar v)(\alpha z_1^2+\beta z_2^2)-(E+a+a'+v+w)(\alpha z_1+
\beta z_2) &= 0\nonumber\\
(u-\bar u)(\alpha z_1^{L-1}+\beta z_2^{L-1})-(E+b+b'+u+s) (\alpha
z_1^L+\beta z_2^L)&= 0.
\end{align}
Defining
\begin{align}\label{21}
\delta a&:=a+a'-(u+s),\nonumber\\
\delta b&:=b+b'-(v+w),
\end{align}
and using (\ref{19}) to eliminate $E$, one arrives at
\begin{align}\label{22}
[(u-\bar u)+z_1\delta a][(v-\bar v)z_2^{L+1}+z_2^L\delta b]&-
[(u-\bar u)+z_2\delta a]\nonumber\\
&\times [(v-\bar v)z_1^{L+1}+z_1^L\delta b]=0.
\end{align}
This is the same as equation (15) in \cite{MA}, with $u$ and $v$
replaced by $u-\bar u$ and $v-\bar v$, respectively. The
qualitative difference between (15) in \cite{MA} and (\ref{22}) is
that $u$ and $v$ are nonnegative, whereas $u-\bar u$ and $v-\bar
v$ may be negative. Defining
\begin{align}\label{23}
Z_j&:= z_j\sqrt{\left\vert{{v-\bar v}\over{u-\bar
u}}\right\vert},\nonumber\\
A&:={\rm sgn}(u-\bar u){{\delta a}\over{\sqrt{|(v-\bar v)(u-\bar
u)|}}},\nonumber\\
B&:={\rm sgn}(v-\bar v){{\delta b}\over{\sqrt{|(v-\bar v)(u-\bar
u)|}}},
\end{align}
(\ref{22}) is simplified to
\begin{equation}\label{24}
Z_2^{L+1}(1+A\; Z_1)(1+B/Z_2)-Z_1^{L+1}(1+A\; Z_2)(1+B/Z_1)=0.
\end{equation}
Using (\ref{19}), it is seen that
\begin{equation}\label{25}
z_1z_2={{u-\bar u}\over{v-\bar v}},
\end{equation}
or
\begin{equation}\label{26}
Z_1Z_2={\rm sgn}[(u-\bar u)(v-\bar v)].
\end{equation}
The eigenvalue $E$ is also written as
\begin{align}\label{27}
E&=-(v+w+u+s)+\sqrt{|(u-\bar u)(v-\bar v)|}[Z{\rm sgn}(v-\bar v)+
Z^{-1}{\rm sgn}(u-\bar u)]\nonumber\\
&=-(v+w+u+s)+{\rm sgn}(v-\bar v)\sqrt{|(u-\bar u)(v-\bar
v)|}(Z_1+Z_2)\nonumber\\
&=-(v+w+u+s)+{\rm sgn}(u-\bar u)\sqrt{|(u-\bar u)(v-\bar
v)|}(Z_1^{-1}+ Z_2^{-1}).
\end{align}
Let's have a closer look at (\ref{24}). Using (\ref{26}), \Ref{24}
is converted to a polynomial equation for $Z_j$, having $2L+2$
roots. For $(u-\bar u)(v-\bar v)>0$, $Z_j=1$ and $Z_j=-1$
obviously satisfy (\ref{24}). For $(u-\bar u)(v-\bar v)<0$,
$Z_j=i$ and $Z_j=-i$ are the trivial solutions of (\ref{24}). But
these solutions lead to
\begin{equation}\label{28}
x_k=z^k(\alpha +\beta k),
\end{equation}
not something like (\ref{18}). And this form for $x_k$, generally
does not satisfy the boundary conditions at $k=1,L$. So the other
(nontrivial) $2L$ roots of (\ref{24}) correspond to the
eigenvalues of $h$.

First consider the case $(u-\bar u)(v-\bar v)>0$. If all of the
roots of $Z_j$ are phases, then
\begin{equation}\label{29}
E\leq -(v+w+u+s)+2\sqrt{(u-\bar u)(v-\bar v)}.
\end{equation}
Equality holds if $Z_j={\rm sgn}(v-\bar v)={\rm sgn}(u-\bar u)$.
Normally, this is not a nontrivial root of (\ref{24}). But in the
thermodynamic limit $L\to\infty$, the nontrivial roots of
(\ref{24}) fill the whole unit circle. So the relaxation time for
this case is
\begin{equation}\label{30}
\tau =[u+v+w+s-2\sqrt{(u-\bar u)(v-\bar v)}]^{-1}.
\end{equation}
It is seen that it does not depend on the injection- and
extraction-rates. If, however, some of the solutions of (\ref{24})
are not phases, then the situation is different. Let $Z_1=Z$ be a
root of (\ref{24}) with $|Z|>1$. In the thermodynamic limit
$L\to\infty$, (\ref{24}) is turned to
\begin{equation}\label{31}
\left(1+{A\over Z}\right)\left(1+{B\over Z}\right)=0,
\end{equation}
which has the solutions
\begin{equation}\label{32}
Z=-A,-B.
\end{equation}
But note that we were seeking solutions with moduli greater than
one. This shows that there is such a solution provided $|A|>1$ or
$|B|>1$. If both hold, there exists two solutions with moduli
greater than one. Suppose $|A|>1$. Putting $Z_1=-A$ in (\ref{19}),
one arrives at
\begin{equation}\label{33}
E=-(v+w+u+s)-{\rm sgn}(u-\bar u)\sqrt{(u-\bar u)(v-\bar v)}
(A+A^{-1}).
\end{equation}
If ${\rm sgn}(u-\bar u)A<0$, this value of $E$ violates
(\ref{29}), and the relaxation time is no more obtained from
(\ref{30}). In this case,
\begin{equation}\label{34}
\tau =[v+w+u+s+{\rm sgn}(u-\bar u)\sqrt{(u-\bar u)(v-\bar v)}
(A+A^{-1})]^{-1},
\end{equation}
which is greater than (\ref{30}), and does depend on the
injection- and extraction- rates. This is the dynamic phase
transition. The point at which this occurs is
\begin{equation}\label{35}
\delta a=-\sqrt{(u-\bar u)(v-\bar v)}.
\end{equation}
In terms of the injection- and extraction- rates, the transition
point is
\begin{equation}\label{36}
a+a'=u+s-\sqrt{(u-\bar u)(v-\bar v)}.
\end{equation}

A similar behavior is seen at the transition point
\begin{equation}\label{37}
b+b'=v+w-\sqrt{(u-\bar u)(v-\bar v)}.
\end{equation}
If the injection- and extraction rates are less than this, then we
have
\begin{equation}\label{38}
\tau =[v+w+u+s+{\rm sgn}(u-\bar u)\sqrt{(u-\bar u)(v-\bar v)}
(B+B^{-1})]^{-1},
\end{equation}

These sound quite similar to the results of \cite{MA}. But there
is a difference. In the models studied in \cite{MA}, either $A$
could be less than $-1$ or $B$, and it was impossible that both be
less than one. The reason is that there $s=w=\bar u=\bar v=0$, and
this means that in the physical region (where all of the rates are
nonnegative), either the left--hand side of (\ref{36}) is always
greater than the right--hand side of (\ref{36}), or the left--hand
side of (\ref{37}) is always greater than the right--hand side of
(\ref{37}), since one of the right--hand sides is nonpositive. But
it is not the case in the present model. Defining
\begin{align}\label{39}
{\cal A}_1&:=u+s-\sqrt{(u-\bar u)(v-\bar v)}=\bar u+\bar
s-\sqrt{(\bar u-u)(\bar v-v)},\nonumber\\
{\cal B}_1&:=v+w-\sqrt{(u-\bar u)(v-\bar v)}=\bar v+\bar
w-\sqrt{(\bar u-u)(\bar v-v)},
\end{align}
it is seen that at least one of ${\cal A}_1$ or ${\cal B}_1$ are
positive (apart from the special case $u=v=\bar s=\bar w$ and
$\bar u=\bar v=s=w=0$, where both of them are zero). But it is
also possible that both of them are positive. In general, there
may be three phases:
\begin{equation}\label{40}
\tau =
\begin{cases}
[v+w+a+a'+(u-\bar u)(v-\bar v)(a+a'-u-s)^{-1}]^{-1},&\text{ region
I}\cr [u+s+b+b'+(u-\bar u)(v-\bar v)(b+b'-v-w)^{-1}]^{-1},&\text{
region II}\cr [v+w+u+s-2\sqrt{(u-\bar u)(v-\bar v)}]^{-1},&\text{
otherwise}
\end{cases}
\end{equation}
Regions I and II are defined as
\begin{equation}\label{41}
a+a'<{\cal A}_1,\quad a+a'-b-b'<{\cal A}_1-{\cal
B}_1,\qquad\text{for region I},
\end{equation}
\begin{equation}\label{42}
b+b'<{\cal B}_1,\quad a+a'-b-b'>{\cal A}_1-{\cal
B}_1,\qquad\text{for region II}.
\end{equation}
So the system may have two phases or three phases, depending on
whether only one of ${\cal A}_1$ and ${\cal B}_1$ are positive or
both of them are positive. In the special case mentioned above,
the system has only one phase. This is the same Glauber model at
zero temperature with diffusion, studied in \cite{MA}.

Next consider the case $(u-\bar u)(v-\bar v)<0$. If all of the
solutions to (\ref{24}) are phases, then (\ref{27}) shows that
\begin{equation}\label{43}
\Re(E)=-(v+w+u+s),
\end{equation}
and from that,
\begin{equation}\label{44}
\tau =(v+w+u+s)^{-1}.
\end{equation}
So in this case the relaxation time does not depend on the
injection- and extraction- rates. Moreover, the eigenvalues of $h$
are complex not real. This means that the relaxation of the system
toward its stationary state is oscillatory.

Now suppose that there exists solutions for (\ref{24}) that are
not phases. Let $|Z_1|>1>|Z_2|$. At the thermodynamic limit, and
using $Z_2=-Z_1^{-1}$, (\ref{24}) is turned to
\begin{equation}\label{45}
\left(1-{A\over Z_1}\right)\left(1+{B\over Z_1}\right)=0.
\end{equation}
The solution to this is
\begin{equation}\label{46}
Z_1=A,-B.
\end{equation}
It is obvious that to have non--phase roots, either $|A|$ or $|B|$
should be greater than 1. Suppose $|A|>1$. Corresponding to
$Z_1=A$, one obtains
\begin{equation}\label{47}
E=-(v+w+u+s)-{1\over{\delta a}}[(\delta a)^2-|(u-\bar u)(v-\bar
v)|].
\end{equation}
As $|A|>1$, the expression in the bracket is positive. If $\delta
a<0$, then this value of $E$ is greater than the right--hand side
of (\ref{43}). So this value of $E$ determines the relaxation
time. That is,
\begin{align}\label{48}
\tau =&\{ v+w+u+s+(\delta a)^{-1}[(\delta a)^2- |(u-\bar u)(v-\bar
v)|]\}^{-1},\nonumber\\
=&[v+w+a+a'-|(u-\bar u)(v-\bar v)|(a+a'-u-s)^{-1}]^{-1}.
\end{align}
A similar argument shows that for $\delta b< -\sqrt{|(u-\bar
u)(v-\bar v)|}$, the relaxation time is
\begin{equation}\label{49}
\tau =[u+s+b+b'-|(u-\bar u)(v-\bar v)|(b+b'-w-v)^{-1}]^{-1}.
\end{equation}
Finally, if both $|A|$ and $|B|$ are greater than 1, then the
larger of (\ref{48}) and (\ref{49}) is the relaxation time.
Defining ${\cal A}_2$ and ${\cal B}_2$ similar to (\ref{39}):
\begin{align}\label{50}
{\cal A}_2&:=u+s-\sqrt{(u-\bar u)(\bar v-v)}=\bar u+\bar
s-\sqrt{(\bar u-u)(v-\bar v)},\nonumber\\
{\cal B}_2&:=v+w-\sqrt{(\bar u-u)(v-\bar v)}=\bar v+\bar
w-\sqrt{(u-\bar u)(\bar v-v)},
\end{align}
one arrives at
\begin{equation}\label{51}
\tau =
\begin{cases}
[v+w+a+a'+(u-\bar u)(v-\bar v)(a+a'-u-s)^{-1}]^{-1},&\text{ region
I}\cr [u+s+b+b'+(u-\bar u)(v-\bar v)(b+b'-v-w)^{-1}]^{-1},&\text{
region II}\cr (v+w+u+s)^{-1},&\text{ otherwise}
\end{cases}
\end{equation}
where the definitions of the regions are the same as (\ref{41})
and (\ref{42}), with ${\cal A}_1$ and ${\cal B}_1$ replaced by
${\cal A}_2$ and ${\cal B}_2$, respectively. Note that at least
one of ${\cal A}_2$ and ${\cal B}_2$ is positive (apart from the
special case $\bar u=\bar v=s=w$ and $u=v=\bar s=\bar w=0$, where
both of them are zero), but it may be that both are positive. If
only one of them is positive, the system has two phases. If both
are positive, it has three phases. In the special case mentioned
above, the system has only one phase.

One can combine (\ref{40}) and (\ref{51}) in a single relation.
First, note that (\ref{39}) and (\ref{50}) can be combined as
\begin{align}\label{50a}
{\cal A}&:=u+s-\sqrt{|(u-\bar u)(v-\bar v)|}=\bar u+\bar
s-\sqrt{|(u-\bar u)(v-\bar v)|},\nonumber\\
{\cal B}&:=v+w-\sqrt{|(u-\bar u)(v-\bar v)|}=\bar v+\bar
w-\sqrt{|(u-\bar u)(v-\bar v)|}.
\end{align}
Then we can write
\begin{equation}\label{52}
\tau =
\begin{cases}
[v+w+a+a'+(u-\bar u)(v-\bar v)(a+a'-u-s)^{-1}]^{-1},&\text{ region
I}\cr [u+s+b+b'+(u-\bar u)(v-\bar v)(b+b'-v-w)^{-1}]^{-1},&\text{
region II}\cr \{v+w+u+s-2\Re[\sqrt{(u-\bar u)(v-\bar
v)}]\}^{-1},&\text{ otherwise}
\end{cases}
\end{equation}
where the definitions of the regions are the same as \Ref{41} and
\Ref{42}, with ${\cal A}_1$ and ${\cal B}_2$ replaced with ${\cal
A}$ and ${\cal B}$, respectively.

\vskip 2\baselineskip\noindent{\bf Acknowledgement}

\noindent The authors would like to thank Institute for Studies in
Theoretical Physics and Mathematics for partial support.

\end{document}